\documentclass[amssymb,amsmath,preprint,superscriptaddress,nofootinbib]{revtex4-1}
\usepackage[colorlinks]{hyperref}
\usepackage{graphicx}

\usepackage{gensymb}

\begin{document}
\title{Thermal leptogenesis in the presence of helical hypermagnetic fields}

\author{S. Safari}
\email{sa.safari@mail.sbu.ac.ir}
\affiliation{Department of Physics, Shahid Beheshti University, Tehran, Iran}

\author{M. Dehpour}
\email{m.dehpour@cern.ch}
\affiliation{Faculty of Mathematics and Physics, Charles University, Prague, Czechia}

\author{S. Abbaslu}
\email{s-abbaslu@ipm.ir}
\affiliation{School of Physics, Institute for Research in Fundamental Sciences (IPM), Tehran, Iran}
\affiliation{Department of Physics, Shahid Beheshti University, Tehran, Iran}

\begin{abstract}
    One of the major challenges in particle physics and cosmology is understanding why there is an asymmetry between matter and antimatter in the Universe. One possible explanation for this phenomenon is thermal leptogenesis, which involves the addition of at least two right-handed neutrinos (RHNs) to the standard model. Another possible explanation is baryogenesis through the hypermagnetic fields which involves the ${\rm U}_Y(1)$ anomaly and helical hypermagnetic fields in the early Universe. 
    In this paper, after reviewing the thermal leptogenesis and baryogenesis through the ${\rm U}_Y(1)$ anomaly, we investigate the simplest model that combines these two scenarios and explore the parameter space for optimal results. 
    Our results show that the combined scenario permits a specific region of parameter space that is not covered by either one separately. In fact, the minimum required mass scale of the RHN and strength of initial hypermagnetic helicity are reduced by one order of magnitude in our model. 
    Moreover, we find that in the combined scenario, leptogenesis and baryogenesis through the ${\rm U}_Y(1)$ anomaly can either amplify or reduce the effect of each other, {\it i.e.}, the generated asymmetry, depending on the sign of the helical hypermagnetic fields.
    Finally, we show the surprising result that a drastic amplification can occur even when the initial abundance of RHN is its equilibrium value for leptogenesis.
\end{abstract}

\maketitle

\section{Introduction}
\label{sec:intro}
Observations have confirmed that the Universe is predominantly composed of matter, rather than antimatter, which is known as the baryon asymmetry of the Universe (BAU). 
The BAU can be defined as the ratio of net baryon number to entropy densities. The value of the baryon asymmetry has been determined via various observations: Big Bang Nucleosynthesis (BBN), Cosmic Microwave Background (CMB), and Large-Scale Structure (LSS)  observations. The results, at 95\% confidence level, are presented as \cite{Simha:2008zj}
\begin{align}
    Y^{\rm obs}_{B} \equiv \frac{n_B - \overline{n}_{B}}{s} = (8.73 \pm 0.35) \times 10^{-11}.
    \label{eq:YBobs}
\end{align}
The origin of BAU is a longstanding problem in particle physics. To explain the BAU, any proposed mechanism of baryogenesis must satisfy the three necessary Sakharov conditions \cite{Sakharov:1967dj}: (i) violation of baryon number, (ii) violation of C and CP symmetry, and (iii) departure from thermal equilibrium. There is a wide range of scenarios within and beyond the standard model of particle physics which attempt to explain this asymmetry, as discussed in various studies (see, for example,  \cite{Elor:2022hpa,DiBari:2021fhs} and references therein).

Leptogenesis is one possible scenarios to explain BAU: By extending the standard model with RHNs, the origin of small neutrino masses can be explained through the type-I seesaw mechanism, while simultaneously addressing the BAU through leptogenesis. The concept of leptogenesis was first introduced by Fukujita and Yanagida \cite{Fukugita:1986hr} and is also known as thermal, standard, or vanilla leptogenesis. However, this scenario has some drawbacks, such as an initial condition problem\footnote{The initial condition problem refers to the challenge of generating final asymmetry fully independent of the initial condition.} \cite{Bertuzzo:2010et} and a large required RHN mass \cite{Davidson:2002qv}. A large RHN mass makes the model phenomenologically non-testable because the energy scale becomes too high to be accessible in laboratory experiments \cite{Dasgupta:2021ies}. This bound also could conflict with supersymmetric models because of the overproduction of gravitino~\cite{Kawasaki:2008qe,Rychkov:2007uq,Kawasaki:1994af,Khlopov:1984pf,Weinberg:1982zq}.
The standard leptogenesis is also referred to as vanilla leptogenesis since the effect of lepton flavor is neglected. In contrast, flavor leptogenesis takes into account the effect of lepton flavors \cite{Nardi:2006fx,Abada:2006fw}, aiming to address the initial condition problem \cite{Bertuzzo:2010et}. Resonant leptogenesis \cite{Pilaftsis:2003gt} and the Akhmedov, Rubakov and Smirnov (ARS) leptogenesis \cite{Akhmedov:1998qx,Asaka:2005pn}, are the most well-known alternative scenarios which try to reduce the required RHN masses. 
Utilizing type-II \cite{Antusch:2004xy} or type-III \cite{Albright:2003xb} seesaw mechanisms, instead of type-I, in thermal leptogenesis can also modify valid regions of the parameter space. Also considering new decay channels, such as an electromagnetic decay channel through a possible magnetic moment of neutrino, can help reach low-scale leptogenesis \cite{Bell:2008fm}.
Leptogenesis in the context of some nonstandard cosmologies has also been investigated \cite{Dehpour:2023dfo,Dehpour:2023wyy,Chen:2019etb,Dutta:2018zkg,PhysRevD.90.064050}.

Observations reveal that there exist long-range magnetic fields throughout the Universe, with amplitudes ranging from $10^{-17}{\rm G}$ in the intergalactic medium to $10^{-9}{\rm G}$ in the intragalactic medium, and with correlation lengths of approximately $\lambda_0 \simeq 10^{-6}\ \rm Mpc$, which are obtained from the measurements of CMB \cite{Planck:2015zrl} and gamma rays from blazars \cite{Ando:2010rb,Tavecchio_2010,Neronov_2010,Essey:2010nd,Chen:2014rsa}. Mechanisms responsible for generating the cosmic magnetic field are known as magnetogenesis, which can be divided into two categories: the early Universe cosmological models \cite{Subramanian:2015lua,Kandus:2010nw} and the astrophysical models \cite{Subramanian:2015lua,wielebinski_cosmic_2005}. 
The generation and amplification of the hypermagnetic fields in the symmetric phase through the chiral anomaly, which is classified in the former category, has received significant attention \cite{Joyce:1997uy,Giovannini:1997eg,Tashiro_2012,Giovannini:2013oga,Giovannini:2015aea,RostamZadeh:2015xnd,RostamZadeh:2016bgf,RostamZadeh:2018zxj,Abbaslu:2020xfn,Abbaslu:2021mkt,Abbaslu:2021zop,Abbaslu:2019yiy}. In these scenarios, due to the $\rm U(1)_Y$ anomaly, a nonzero seed of hypermagnetic fields can be amplified through the chiral magnetic effect (CME). Going through the electroweak phase transition, these fields are partially converted into ordinary Maxwellian magnetic fields \cite{Kamada:2016cnb}. The generation of strong hypermagnetic fields and the required baryon asymmetry from zero initial values have also been addressed by considering the chiral vortical effect \cite{Tashiro_2012,Giovannini:1997eg,Giovannini:2015aea,Giovannini:2013oga,Abbaslu:2020xfn}.
In general, there are three main categories of the relationship between the generation and evolution of hypermagnetic field, and BAU: (i) producing the BAU from the hypermagnetic field \cite{Giovannini:1997eg,Giovannini:1997gp,Giovannini:1999wv,Vilkovisky:1999ds,Bamba:2006km,Bamba:2007hf,Dvornikov:2011ey,Dvornikov:2012rk,Fujita:2016igl}, (ii) the reverse of this process \cite{Giovannini:1997eg,Long:2016uez,Joyce:1997uy}, and (iii) generating both of them simultaneously, {\it e.g.} using the chiral vortical effect \cite{Abbaslu:2020xfn,Abbaslu:2021zop,Giovannini:2013oga,Giovannini:2015aea}.

It is known that, in the presence of external hypermagnetic fields, the ${\rm U}_Y(1)$ anomaly violates the conservation of total lepton ($L$) and baryon ($B$) numbers, while their difference, $B-L$, remains constant \cite{Adler:1969gk,Bell:1969ts,PhysRevD.42.3344}. In the context of thermal leptogenesis, the decay of RHN(s) leads to the violation of $L$. Subsequently, the weak sphalerons can convert some of the generated $L$ into $B$, while preserving $B-L$. 

Although baryogenesis through the ${\rm U}_Y(1)$ anomaly and the thermal leptogenesis have been studied separately in the literature, to the best of our knowledge, the combined scenario has not been presented before. 
In this study, we focus on baryogenesis by combining baryogenesis through the ${\rm U}_Y(1)$ anomaly and thermal leptogenesis, the latter being based on extending the standard model with three RHNs. 
In our model, the evolution equations for the left-handed leptons and Higgs asymmetry, within the baryogenesis through the ${\rm U}_Y(1)$ anomaly scenario, each gain an additional source term resulting from the decay of the lightest RHN. Meanwhile, the evolution equations for the asymmetries of right-handed leptons and all quarks remain unchanged. In each lepton and quark generation sector, the right-handed and left-handed fermions are connected through the chirality-flip processes. Meanwhile, the weak sphaleron interactions, which involve only left-handed fermions, can interconnect all quarks and leptons more effectively when all of the mentioned processes are in equilibrium. For clarity, in this study we have allowed all chirality-flip processes, as well as weak and strong sphaleron processes to be out of thermal equilibrium.

As we shall show, in the combined scenario, lower values of RHNs masses and initial hypermagnetic helicity can generate the desired BAU, as compared to the cases in which these two scenarios are considered separately. Moreover, in the combined scenario, the two mechanisms of baryogenesis through the ${\rm U}_Y(1)$ anomaly and leptogenesis can either dramatically amplify or reduce the effect of each other, depending on the sign of the helical hypermagnetic fields. More importantly, we show that this dramatic increase occurs even if the initial abundance of the RHN is its equilibrium value. In fact, all leptogenesis scenarios can produce the desired BAU, provided that the initial abundance of the lightest RHN deviates from its equilibrium value. However, in the temperature range of our interest, it is hard to justify departure from the equilibrium abundance of the RHNs. Thus, our combined model circumvents this difficulty, as well.

This paper is structured as follows: In Sect.~\ref{sec:lep}, we briefly review the thermal leptogenesis mechanism for producing BAU. In Sect.~\ref{sec:magnetic} we write the anomalous Maxwell equations and explain the baryogenesis mechanism through the helical hypermagnetic fields. In Sect.~\ref{sec:leptogenesis-magnetic}, we derive the evolution equations of the asymmetries by taking into account both thermal leptogenesis and baryogenesis through the ${\rm U}_Y(1)$ anomaly. In Sect.~\ref{sec:res}, we numerically solve the set of coupled differential equations obtained in Sect.~\ref{sec:leptogenesis-magnetic}. Finally, in Sect.~\ref{sec:conclu}, we discuss our results.

\section{Baryogenesis through thermal leptogenesis}
\label{sec:lep}
Thermal leptogenesis is based on the extension of the Standard Model by the addition of at least two RHNs\footnote{According to Eq.\ \ref{eq:CP-parameter}, one RHN cannot violate CP.}, which can interact with standard model particles just through Yukawa interactions and gravity. These sterile and heavy particles can be created through thermal mechanisms in the early Universe.

If we assume the existence of three RHNs, with only the lightest one participating in the Yukawa interaction, we can describe the Yukawa matrix in the context of the type-I seesaw mechanism using the Casas-Ibarra parametrization \cite{Casas:2001sr},
\begin{align}
    y = \frac{i}{v} U m^{1/2} R M^{1/2},
    \label{eq:Casas-Ibarra}
\end{align}
where $v=174\ \rm GeV$ denotes the Higgs expectation value, $m$ is the diagonal mass matrix of light neutrinos, $M$ is the diagonal mass matrix of RHNs, $U$ is the unitary neutrino mixing matrix known as PMNS matrix (Pontecorvo-Maki-Nakagawa-Sakata matrix), and $R$ is a complex orthogonal matrix.

The PMNS can be decomposed as follows \cite{Chau:1984fp}:
\begin{align}
    U = 
    \begin{pmatrix}
	1&0&0\\0&c_{23}&s_{23}\\0&-s_{23}&c_{23}
    \end{pmatrix}
    \begin{pmatrix}
        c_{13}&0&s_{13}e^{-i\delta}\\0&1&0\\s_{13}e^{i \delta}&0&c_{13}
    \end{pmatrix}
    \begin{pmatrix}
	c_{12}& s_{12} & 0\\-s_{12} & c_{12} & 0 \\ 0 & 0 & 1
    \end{pmatrix},
    \label{eq:PMNS}
\end{align}
where $\delta$ is the Dirac phase\footnote{Note that there are also two phases of Majorana, $\alpha_{21}$ and $\alpha_{31}$ that can take values between $0$ and $4\pi$. Here we neglected $\alpha_{21}$ and $\alpha_{31}$, as there is no experimental approach to determine these \cite{Esteban:2020cvm}.}, $c_{ij} = \cos \theta_{ij}$ and $s_{ij} = \sin \theta_{ij}$ and $\theta_{ij}$ denote the mixing angles. Moreover, for the final piece of the Yukawa matrix expressed in Eq.\ (\ref{eq:Casas-Ibarra}), the matrix $R$ can be decomposed using $\omega_i = x_i + i y_i$, and expressed in the following form,
\begin{align}
    R=
    \begin{pmatrix}
        1&0&0\\0&c_{\omega_1}&s_{\omega_1}\\0&-s_{\omega_1}&c_{\omega_1}
    \end{pmatrix}
    \begin{pmatrix}
	c_{\omega_2}&0&s_{\omega_2}\\0&1&0\\-s_{\omega_2}&0&c_{\omega_2}
    \end{pmatrix}
    \begin{pmatrix}
	c_{\omega_3}&s_{\omega_3}&0\\-s_{\omega_3}&c_{\omega_3}&0\\0&0&1
    \end{pmatrix},
    \label{eq:R-matrix}
\end{align}
where $c_{\omega_i} = \cos \omega_i$ and $s_{\omega_i} = \sin \omega_i$.

The participation of the lightest RHN in the Yukawa interaction makes its decay capable of CP violation and producing lepton number asymmetry. More precisely, the reactions involving $N_1$ can be described as follows:
\begin{align}
    N_1 \rightleftarrows \bar{\varphi} l,
    \label{eq:decay}\\
    N_1 \rightleftarrows \varphi \bar{l},
    \label{eq:decay-anti}
\end{align}
where $\varphi$ and $l$ denote the Higgs and left-handed lepton doublets, respectively. To quantify this scenario, one can first calculate the tree-level decay rates \cite{Davidson:2008bu},
\begin{align}
    \Gamma_1= \overline{\Gamma}_1 = \frac{M_1}{16 \pi} (yy^{\dagger})_{11}.
    \label{eq:decay-rate}
\end{align}
where $\Gamma_1$ denotes $\Gamma(N_1 \to \bar{\varphi} l)$ and $\overline{\Gamma}_1$ denotes $\Gamma(N_1 \to \varphi \bar{l})$. In order to gain a better understanding of how particles behave in different conditions and to explore the dynamical relationship between particle energies and temperature, we take the thermal average of decay rates to obtain \cite{Kolb:1979qa}
\begin{align}
    \langle\Gamma_{1}\rangle = \langle\overline{\Gamma}_1\rangle =  \frac{K_1(z)}{K_2(z)} \frac{M_1}{16 \pi} (yy^{\dagger})_{11},
    \label{eq:averaged-decay-rate}
\end{align}
where $\langle \dots \rangle$ denotes thermal averaging by the Maxwell-Boltzmann distribution, where $z\equiv M_1/T$ is a dimensionless parameter, and $K_n(z)$ is the modified Bessel function of the second kind.

Due to the expansion of the Universe, the temperature decreases and eventually reaches values lower than $M_1$. As a result, the reactions expressed by Eqs.~(\ref{eq:decay}) and (\ref{eq:decay-anti}) only proceed in one direction, from left to right, leading to decays. If their decay rates are not equal, then CP violation takes place.
We can then proceed to introduce a CP violation parameter which is adjusted to the total decay rate and can be expressed as
\begin{align}
    \epsilon_1  \equiv \frac{\Gamma_1 - \overline{\Gamma}_1}{\Gamma_1 + \overline{\Gamma}_1}.
\end{align}
The CP violation parameter is nonzero only if loop corrections are taken into account \cite{Davidson:2008bu}
\begin{align}
    \epsilon_1 = \sum_{k\neq1} \frac{1}{8\pi} \frac{\Im \left(yy^{\dagger}\right)_{1k}^2}{\left(yy^{\dagger}\right)_{11}} \left[ f\left(\frac{M_k^2}{M_1^2}\right) + \frac{M_1 M_k}{M_1^2 - M_k^2}\right],
    \label{eq:CP-parameter}
\end{align}
where function $f(x)$ is given by
\begin{align}
    f(x) = \sqrt{x} \left[1-\left(1+x\right)\ln\left(\frac{1+x}{x}\right)\right].
\end{align}

The evolution of the lepton asymmetries and the number densities of RHNs can be obtained using the Boltzmann equations in the context of the Friedmann-Lemaitre-Robertson-Walker (FLRW) Universe. Hence, we can obtain \cite{Buchmuller:2004nz,Davidson:2008bu}:
\begin{align}
    \frac{dY_{N_1}}{dz} &= - D_1 \left( Y_{N_1} - Y_{N_1}^{\rm  eq} \right), 
    \label{eq:YN1}\\
    \frac{dY_{B-L}}{dz} &= -\epsilon_1 D_1 \left( Y_{N_1} - Y_{N_1}^{\rm  eq} \right) - W_1 Y_{B-L}
    \label{eq:YBL},
\end{align}
where $Y_{N_1}\equiv n_{N_1}/s$ is normalized RHN number density, $Y_{B-L}$ is the $B-L$ asymmetry, and  $s=2 \pi^2 g_{\star} T^3/45$ is entropy density. Note that $g_{\star}=106.75$ is the effective number of relativistic degrees of freedom \cite{Husdal:2016haj}.
Moreover, the decay parameter $D_1$ and the washout parameter $W_1$ are defined as \cite{Buchmuller:2004nz}
\begin{align}
    D_1 \equiv  \frac{2 \langle\Gamma_{1}\rangle}{H z}, \quad
    W_1 \equiv \frac{1}{2} \frac{Y_{N_1}^{\rm  eq}}{Y_{l}^{\rm  eq}} D_1,
\end{align}
where $H$ represents the Hubble parameter, which takes the following form in the radiation-dominated phase \cite{Kolb:1990vq}
\begin{align}
    H = \frac{1.66}{M_{\rm Pl}} g_{\star}^{1/2} \left(\frac{M_1}{z}\right)^2,
\end{align}
where the Planck mass $M_{\rm Pl} = 1.22 \times 10^{19}\ \rm GeV$. Furthermore, $Y_{N_1}^{\rm  eq}$ and $Y_{l}^{\rm  eq}$ denote the equilibrium value of the normalized number density of RHN and lepton given by \cite{Kolb:1990vq}
\begin{align}
    Y_{N_1}^{\rm  eq} = \frac{45}{4 \pi^4} \frac{g_{N_1}}{g_{\star}} z^2 K_2(z), \quad 
    Y_{l}^{\rm  eq} \simeq \frac{45}{4 \pi^4} \frac{g_{l}}{g_{\star}} \frac{3}{2} \zeta(3) \simeq 0.0039,
    \label{eq:YEq}
\end{align}
where $g_{N_1}=g_{l}=2$ are degree of freedoms, and $\zeta(s)$ is the Riemann zeta function.

In this mechanism, the decay of the RHN first leads to the generation of a left-handed lepton asymmetry, denoted as $Y_{B-L}$. Subsequently, the generated $Y_{B-L}$ can convert into the baryon asymmetry $Y_B$ through the weak sphaleron process. As the Universe expands, the Hubble parameter decreases with the square of the temperature, while the rates of both perturbative and nonperturbative interactions decrease with temperature. As time passes, reactions which were initially insignificant become increasingly effective and eventually attain thermal equilibrium. Reactions characterized by higher rates reach equilibrium more quickly. Upon taking into account the hypercharge neutrality condition, the weak and strong sphaleron processes, as well as all chirality-flip processes in the thermal equilibrium, the baryon asymmetry is obtained as \cite{Chen:2007fv} 
\begin{align}
    Y_{B}^{\rm Lepto.}(z^{\rm eq}) = \frac{28}{79} Y_{B-L}(z^{\rm eq}),
\end{align}
where $z^{\rm eq}$ denotes the time when all perturbative and nonperturbative processes are in thermal equilibrium.

In summary, baryogenesis through thermal leptogenesis satisfies all three Sakharov conditions simultaneously: the decay of RHN violates CP\footnote{The weak interactions of the standard model violate C maximally.} due to loop corrections, this process occurs out of equilibrium and supplies the lepton number violation which is partially converted to the baryon number violation by weak sphaleron processes.

\section{Baryogenesis through helical hypermagnetic fields}
\label{sec:magnetic}
In this section, we briefly review the model for baryogenesis through the ${\rm U}_Y(1)$ anomaly in the symmetric phase of the early Universe. As is well known, the lepton and quark current conservations are violated due to the Abelian and non-Abelian anomalies \cite{Adler:1969gk,Bell:1969ts,tHooft:1976rip,Long:2013tha}. These anomaly relations are given by \cite{Adler:1969gk,Bell:1969ts,tHooft:1976rip,Long:2013tha}:
\begin{align}
    \nabla_{\mu}j_{{q}^i_L}^{\mu}&=(N_{c}N_{w}Q_{q}^{2})\frac{g'^{2}}{16\pi^2}Y_{\mu\nu}\tilde{Y}^{\mu\nu}+\frac{1}{2}(N_{c})\frac{g^{2}}{16\pi^2}W_{\mu\nu}^{a}\tilde{W}^{a\, \mu\nu}+\frac{1}{2}(N_{w})\frac{g_{s}^{2}}{16\pi^2}G_{\mu\nu}^{A}\tilde{G}^{A\, \mu\nu},
    \label{eq:er5}\\
    \nabla_{\mu}j_{{u}^{i}_R}^{\mu}&=-(N_{c}Q_{u}^{2})\frac{g'^{2}}{16\pi^2}Y_{\mu\nu}\tilde{Y}^{\mu\nu}-\frac{1}{2}\frac{g_{s}^{2}}{16\pi^2}G_{\mu\nu}^{A}\tilde{G}^{A\, \mu\nu},
    \label{eq:er4}\\
    \nabla_{\mu}j_{{d}^{i}_R}^{\mu}&=-(N_{c}Q_{d}^{2})\frac{g'^{2}}{16\pi^2}Y_{\mu\nu}\tilde{Y}^{\mu\nu}-\frac{1}{2}\frac{g_{s}^{2}}{16\pi^2}G_{\mu\nu}^{A}\tilde{G}^{A\, \mu\nu},
    \label{eq:er3}\\
    \nabla_{\mu}j_{{l}^i_L}^{\mu}&=(N_{w}Q_{l}^{2})\frac{g'^{2}}{16\pi^2}Y_{\mu\nu}\tilde{Y}^{\mu\nu}+\frac{1}{2}\frac{g^{2}}{16\pi^2}W_{\mu\nu}^{a}\tilde{W}^{a\, \mu\nu},
    \label{eq:er2}\\
    \nabla_{\mu} j_{{e}^i_R}^{\mu}&=-(Q_{e}^{2})\frac{g'^{2}}{16 \pi^2}Y_{\mu\nu}\tilde{Y}^{\mu\nu}.
    \label{eq:er1}
\end{align}
Here, $\nabla_{\mu}$ refers to the covariant derivative with respect to the FLRW metric. The index $i$ denotes the generation.
The right-handed singlet and left-handed doublet lepton currents are represented by $j_{{e}^i_R}^{\mu}$ and $ j_{{l}^i_L}^{\mu}=j_{e_L^i}^{\mu}+j_{\nu_L^i}^{\mu}$, respectively. Similarly, the right-handed singlet down and up quark currents are represented by $ j_{{d}^i_R}^{\mu}$, $ j_{{u}^i_R}^{\mu}$, and the left-handed doublet quark current is represented by $ j_{{q}^i_L}^{\mu}=j_{u_L^i}^{\mu}+j_{d_L^i}^{\mu}$. Also, $\rm N_{c}=3$ and $\rm N_{w}=2$ represent the ranks of the non-Abelian $\rm SU(3)$ and $\rm SU_{L}(2)$ gauge groups, respectively. More, $G_{\mu \nu}^{A}$, $W_{\mu \nu}^{a}$, and $Y_{\mu \nu}$ are the field strength tensors of the ${\rm SU}(3)$, ${\rm SU}_{\textrm{L}}(2)$, and ${\rm U}_{\textrm{Y}}(1)$ gauge groups with relevant coupling constant $g_{\rm s}$, $g$, and $g^{\prime}$, respectively. The fine structure constants for these groups are $\alpha_{\rm s}=\frac{g_{s}^2}{4\pi}$, $\alpha_{\rm w}=\frac{g^2}{4\pi}$, and $\alpha_{\rm Y}=\frac{g'^2}{4\pi}$. Moreover, the dual gauge field strength tensors are represented by
${\tilde X}^{\mu\nu} \equiv (1/2) \epsilon^{\mu\nu\rho\sigma} X_{\rho \sigma}/\sqrt{-{\rm det} (g_{\mu\nu})}= (1/2) a^{-3}(t) \epsilon^{\mu\nu\rho\sigma} X_{\rho \sigma}$,and $\epsilon^{\mu\nu\rho\sigma}$ 
is the antisymmetric Levi-Civita symbol specified by $\epsilon^{0123}=-\epsilon_{0123}=1$. Here, $a^{-3}(t)$ is a scaling factor related to the expansion of the Universe. The relevant hypercharges are
\begin{align}
    Q_{l}=-\frac{1}{2}, \quad Q_{e}=-1, \quad Q_{q}=\frac{1}{6}, \quad Q_{u}=\frac{2}{3}, \quad Q_{d}=-\frac{1}{3}.
    \label{eq:eqds1}
\end{align}

The terms proportional to $W_{\mu\nu}^{A}\tilde{W}^{A\, \mu\nu}$ and $G_{\mu\nu}^{A}\tilde{G}^{A\, \mu\nu}$ on the r.h.s of the above equations, at high temperatures, induce the so-called weak and strong sphaleron processes, respectively.
At high temperatures, above the electroweak scale, the weak and strong sphaleron rates are estimated by the numerical simulations as $\Gamma_{\rm{w}}\simeq25\alpha_w^5 T$ \cite{Moore:1997sn,Bodeker:1999gx} and $\Gamma_{\rm{s}}\simeq100\alpha_s^5 T$ \cite{Moore:1997im}.
The weak sphaleron processes can affect only the left-handed quark and lepton numbers. This type of anomaly leaves the right-handed fermions unchanged unless their relevant chirality-flip processes are also active \cite{Long:2013tha}. Meanwhile, the strong sphaleron processes involve only the chiral quarks and affect their chiralities but do not change the baryon number \cite{Long:2013tha,McLerran:1990de}.

Since the vacuum structure of the $U(1)$ gauge theory is trivial, the Abelian anomaly, represented by the first terms on the r.h.s of Eqs.\ (\ref{eq:er5})-(\ref{eq:er1}), has no sphaleron-like structure and violates the chiral quark and lepton numbers only through the time variation of the hypermagnetic helicity. Large electrical conductivity of the electroweak plasma is one of the ever-present sources for the time variation of the hypermagnetic helicity, the expression for which is given by \cite{Fujita:2016igl}:
\begin{align}
    \partial_{t}h =\frac{1}{2}a^3\lim_{V\to \infty} \frac{1}{V}\int_V d^3 x Y_{\mu\nu}\tilde{Y}^{\mu\nu}.
\end{align}
Here, $ h \equiv \lim_{V \to \infty} \frac{\cal H}{V}$ is the hypermagnetic helicity density, with $\cal H$ representing hypermagnetic helicity which is defined as:
\begin{align}
    {\cal H} \equiv \int_V d^3 x \vec{Y}\cdot\vec{B}_{\rm Y}=\int_V d^3 x \epsilon_{ijk} Y_i \partial_j Y_,
\end{align}
where $\vec{Y}$ is the hypercharge vector potential and $\vec{B}_{\rm Y}\equiv {\vec{\nabla}}\times\vec{Y}$.

In addition to the nonperturbative anomalous effects, perturbative chirality-flip processes also play a significant role in the evolution of chiral quark and lepton asymmetries. However, these processes do not directly contribute to lepton and baryon number violations. Instead, they facilitate the erasure of baryon and lepton numbers through weak sphaleron processes by converting right-handed leptons and quarks into their left-handed counterparts, and vice versa. The rate of these processes, which depend on their Yukawa interaction, varies for each fermion. Therefore, during the early stages of the evolution of the Universe, some of these processes are out of thermal equilibrium, which fulfills the third condition outlined by Sakharov. In this study, we consider all chirality-flip processes in the evolution equations and allow them to be out of thermal equilibrium.

Now we write the fermionic asymmetry evolution equations by taking into account all perturbative and nonperturbative processes. After taking the spatial average of Eqs.\ (\ref{eq:er5})-(\ref{eq:er1}) the evolution equations of the quarks, leptons, and Higgs asymmetries are given by \cite{Fujita:2016igl}
\begin{align}
    \frac{\partial Y_{q^i_L}}{\partial z} &= \sqrt{\frac{90}{\pi^2 g_{\star}}} \frac{M_{\rm Pl}^{\prime}}{M_1}\left[-N_{\rm c} N_{\rm w} Q_q^2 \gamma_{\rm y} - N_c \mathcal{S}_{\rm w,sph} - N_{\rm w} \mathcal{S}_{\rm s,sph} - \sum_j \left( \mathcal{S}_{\rm uhu}^{ij} + \mathcal{S}_{\rm dhd}^{ij} \right)\right],
    \label{eq:bol1} \\
    \frac{\partial Y_{u^i_R}}{\partial z} &= \sqrt{\frac{90}{\pi^2 g_{\star}}} \frac{M_{\rm Pl}^{\prime}}{M_1}\left[N_{\rm c} Q_u^2 \gamma_{\rm y} + \mathcal{S}_{\rm s,sph} + \sum_j \mathcal{S}_{\rm uhu}^{ji}\right],
    \label{eq:bol3}\\
    \frac{\partial Y_{d^i_R}}{\partial z} &= \sqrt{\frac{90}{\pi^2 g_{\star}}} \frac{M_{\rm Pl}^{\prime}}{M_1}\left[N_{\rm c} Q_d^2 \gamma_{\rm y} + \mathcal{S}_{\rm s,sph} + \sum_j \mathcal{S}_{\rm dhd}^{ji}\right],
    \label{eq:bol4}\\
    \frac{\partial Y_{l^i_L}}{\partial z} &= \sqrt{\frac{90}{\pi^2 g_{\star}}} \frac{M_{\rm Pl}^{\prime}}{M_1}\left[-N_{\rm w} Q_l^2 \gamma_{\rm y} - \mathcal{S}_{\rm w,sph} - \sum_j \mathcal{S}_{\rm Lhe}^{ij}\right],
    \label{eq:bol2}\\
    \frac{\partial Y_{e^i_R}}{\partial z} &= \sqrt{\frac{90}{\pi^2 g_{\star}}} \frac{M_{\rm Pl}^{\prime}}{M_1}\left[Q_e^2 \gamma_{\rm y} + \sum_{j} \mathcal{S}_{\rm Lhe}^{ji}\right],
    \label{eq:bol5}\\
    \frac{\partial Y_{\varphi}}{\partial z} &= \sqrt{\frac{90}{\pi^2 g_{\star}}} \frac{M_{\rm Pl}^{\prime}}{M_1}\left[\sum_{i,j} \left( \mathcal{S}_{\rm dhd}^{ij} - \mathcal{S}_{\rm uhu}^{ij} + \mathcal{S}_{\rm Lhe}^{ij} \right)\right]. 
    \label{eq:bol6}
\end{align}
Here, $i=1,2,3$ indicates the generation index, 
$M_{\rm Pl}^{\prime} = 2.43 \times 10^{18}\ \rm GeV$ is reduced Planck mass, $Y$'s represent the asymmetries with the subscripts defined analogously to those of Eqs.\ (\ref{eq:er5})-(\ref{eq:er1}), and  $Y_{\varphi}=Y_{\varphi^+}+Y_{\varphi^0}$ denote the asymmetry of the Higgs doublet, with $\varphi^+$ and $\varphi^0$ being the upper and lower Higgs doublet elements.
Moreover, the strong and weak sphaleron processes are represented by
\begin{align}
    \mathcal{S}_{\rm s,sph} &= \gamma_{\rm s}\sum_j \left(Y_{q^j_L} - Y_{u^j_R} - Y_{d^j_R}\right), \\
    \mathcal{S}_{\rm w,sph} &= \gamma_{\rm w}\sum_j \left(Y_{q^j_L}+Y_{l^j_L}\right),
\end{align}
where, $\gamma_{\rm w}\equiv\Gamma_{\rm w}/T \simeq 25 \alpha_{\rm w}^5$ is the dimensionless weak sphaleron rate, and $\gamma_{\rm s}\equiv\Gamma_{\rm s}/T\simeq 100 \alpha_{\rm s}^5$ is the dimensionless strong sphaleron rate.
Also, the source terms $\mathcal{S}_{\rm uhu}^{ij}$, $\mathcal{S}_{\rm dhd}^{ij}$, and $\mathcal{S}_{\rm Lhe}^{ij}$ arise from the Yukawa interactions and are given by
\begin{align}
    \mathcal{S}_{\rm uhu}^{ij} &= \gamma_{u}^{ij} \left(\frac{Y_{q^i_L}}{6} + \frac{Y_{\varphi}}{2} - \frac{Y_{u^j_R}}{3}\right), \\
    \mathcal{S}_{\rm dhd}^{ij} &= \gamma_{d}^{ij} \left(\frac{Y_{q^i_L}}{6} + \frac{Y_{\varphi}}{2} - \frac{Y_{d^j_R}}{3}\right), \\
    \mathcal{S}_{\rm Lhe}^{ij} &= \gamma_{e}^{ij} \left(\frac{Y_{l^i_L}}{2} - \frac{Y_{\varphi}}{2} - Y_{e^j_R}\right).
\end{align}
Here, $\gamma_{u,d,e}^{ij}$ denote the dimensionless Yukawa rates for up-type quarks, down-type quarks, and electron-type leptons with $i,j$ indices. These rates are given by $|y_{u,d,e}^{ij}|^2/8\pi$, where $y_{u,d,e}$ are the corresponding Yukawa coupling matrices, as expressed in Eqs.~(\ref{eq:Casas-Ibarra}), (\ref{eq:PMNS}) and (\ref{eq:R-matrix}), \cite{Fujita:2016igl},
\begin{align}
    &y_{u} \approx \left(\begin{array}{ccc} 1.1 \times 10^{-5} & 0 & 0 \\ 0 & 7.1 \times 10^{-3} & 0 \\ 0&0& 0.94\end{array} \right), 
    \label{eq:yukawa-couplings1}\\
    &y_{d}\approx\left(\begin{array}{ccc} 2.7 \times 10^{-5} & 6.3 \times 10^{-6}  & 2.4 \times 10^{-7}\\ 1.2 \times 10^{-4} & 5.4 \times 10^{-4} & 2.2 \times 10^{-5} \\ 8.3 \times 10^{-5} &9.8 \times 10^{-4}& 2.4 \times 10^{-2} \end{array} \right),
    \label{eq:yukawa-couplings2}\\
    &y_{e}\approx\left(\begin{array}{ccc} 2.8 \times 10^{-6} & 0 & 0 \\ 0 & 5.8 \times 10^{-4} & 0 \\ 0&0& 1.0 \times 10^{-2} \end{array} \right). 
    \label{eq:yukawa-couplings3}
\end{align}

The time derivative of the hypermagnetic helicity, appearing as the source term $\gamma_{\rm y}$ in Eqs.\ (\ref{eq:bol1})-(\ref{eq:bol6}), is given by \cite{Fujita:2016igl}:
\begin{align}\label{gamma y}
    \gamma_{\rm y}\equiv a^{-3} \frac{\alpha_{\rm y}}{2 \pi s}\frac{\dot h}{T}.
\end{align}
Upon using Ampere's law and the generalized Ohm's law, the time evolution of the helicity density is obtained as follows:
\begin{align}
    \dot{h}(t)&=\lim_{V\to \infty} \frac{2}{V} \int_V d^3 x \epsilon_{ijk} \dot{Y}i\partial_j Y_k \notag \\
    &=- a^2 \frac{2}{\sigma} \left\langle\vec{B}_{Y}(t,\vec{x})\cdot \vec{\nabla} \times\vec{B}_Y(t,\vec{x})\right\rangle \notag \\
    &\simeq \mathrm{sgn}(\dot{h}) a^{3} \frac{4\pi }{\sigma} \frac{B_p^2(t)}{\lambda_B(t)},
\end{align}
where $\langle \dots \rangle$ denote spatial averaging, $B_p$ and $\lambda_B$ represent the strength of the physical magnetic field and its correlation length, respectively \cite{Fujita:2016igl}. In this simple form of anomalous equations, the impact of the chiral magnetic effect (CME) is ignored in the Maxwell equations. Consequently, the CME does not appear in the time derivative of the hypermagnetic helicity. It is shown that considering the CME can suppress the growth of the baryon asymmetry in models with a strong hypermagnetic field~\cite{Kamada:2016eeb}. Moreover, the induction of B+L asymmetry from the conversion of the hypermagnetic field to (electro)magnetic field at the electroweak symmetry breaking is ignored~\cite{Kamada:2016cnb}. If this conversion is not completed by the time of the sphaleron freeze-out, the relic baryon asymmetry will be enhanced. For a more complete form of the asymmetry evolution, see Refs.~\cite{Kamada:2016eeb,Kamada:2016cnb}. Taking into account these effects, along with flavor leptogenesis, can be explored in our generalization work.

As mentioned in the Introduction, we do not focus on the origin of the hypermagnetic field in this paper.\footnote{It has been demonstrated that this type of primordial magnetic field can be generated through the chiral vortical effect \cite{Giovannini:1997eg,Tashiro_2012,Giovannini:2013oga} or another mechanism \cite{Subramanian:2015lua,Kandus:2010nw,wielebinski_cosmic_2005,Semikoz:2005ks} at high temperature.} However, their time evolution can be assessed based on their present strength $B_0$ and the correlation length $\lambda_0$ \cite{Fujita:2016igl}. We follow the approach used in Ref. \cite{Fujita:2016igl}, where the hypermagnetic fields at high temperature are estimated in terms of today's magnetic fields strength, $B_0$, and their correlation length, $\lambda_0$. The authors of \cite{Fujita:2016igl} have taken the magnetic field's evolution, influenced by the magnetohydrodynamic, into account to generate the BAU. Since the magnetohydrodynamical effects have the potential to induce the inverse cascade process, the magnetic field's temporal evolution is not strictly adiabatic, where the magnetic field's physical strength diminishes proportionally to $a^{-2}(t)$, with $a(t)$ representing the scale factor. Based on an analytical estimate, they have demonstrated that the observed BAU can be explained, provided that the magnetic field has undergone an inverse cascade process above the electroweak scale. They have explored three distinct situations: (i) the hypermagnetic fields continuously experience inverse cascade starting at their inception, (ii) the hypermagnetic fields initially evolve adiabatically, then enter the inverse cascade regime at a transition temperature $T=T_{\rm TS}$, and (iii) The hypermagnetic fields continuously undergo adiabatic evolution starting at their inception.

Assuming that the present-day magnetic field is maximally helical, the hypermagnetic helicity source term $\gamma_{y}$ in case (ii) above can be estimated as follows \cite{Fujita:2016igl}:
\begin{align}
    \gamma_{\rm y} &\simeq  1.7 \times 10^{-26} \mathcal{C}\left(\frac{B_0}{10^{-14}\ {\rm G}}\right)^2\left(\frac{\lambda_0}{10^{-6}\ {\rm Mpc}}\right)^{-1}\times \left\{\begin{array}{ll}\left(\dfrac{T}{1\ {\rm GeV}}\right)^{4/3} & \quad\text{for} \quad T<T_{\rm TS} \\ \left(\dfrac{T_{\rm TS}}{1\ {\rm GeV}}\right)^{4/3} &\quad  \text{for} \quad T>T_{\rm TS}
    \end{array}\right. . 
    \label{eq:ICgy}
\end{align}
Here, $\mathcal{C} \propto \mathrm{sgn}(\dot{h})$ represents a numerical factor of order unity which accounts for the uncertainty resulting from certain approximations made in this model \cite{Fujita:2016igl}.

Note that we can use the hypercharge neutrality constraint to remove one of the equations presented in  Eqs.\ (\ref{eq:bol1})-(\ref{eq:bol6}). Unlike the $\gamma_{d}$ matrix, $\gamma_{e}$ and $\gamma_{u}$ matrices are diagonal, as can be seen from Eqs.\ (\ref{eq:yukawa-couplings1})-(\ref{eq:yukawa-couplings3}). Therefore, we choose to remove the evolution equation of the third-generation down-like quark by using the hypercharge neutrality constraint, as follows
\begin{align}
    Y_{d_R^3} = \sum_i \left( \frac{1}{2}Y_{q_L^i} + 2 Y_{u_R^i} - \frac{1}{2}Y_{l_L^i} - Y_{e_R^i} \right) + \frac{1}{6} Y_{\varphi}- \sum_{i \neq 3} Y_{d_R^i}.
    \label{eq:hypercharge}
\end{align}

Finally, by solving the above coupled evolution equations simultaneously, the baryon asymmetry can be determined by using the expression
\begin{align}
    Y_B^{\rm U_{Y}(1)} = \frac{1}{3} \sum_i (Y_{q_L^i}+Y_{u_R^i}+Y_{d_R^i}).
\end{align}

\section{Baryogenesis through thermal leptogenesis in the presence of helical hypermagnetic fields}
\label{sec:leptogenesis-magnetic}
This section focuses on the thermal leptogenesis in the presence of background helical hypermagnetic fields. That is, in this scenario, which is a combination of the two scenarios mentioned earlier,  there are two separate sources for the generation of left-handed leptons. The first source arises from the Abelian anomaly in the presence of a nonzero primordial hypermagnetic field, while the second source is attributed to the decay of RHNs. Furthermore, the evolution of the Higgs asymmetry is affected by an additional source term arising from the decay of RHNs.
Therefore, we combine the Boltzmann equations of leptogenesis with the evolution equations of baryogenesis through the ${\rm U}_Y(1)$ anomaly, which was discussed in Sects.~\ref{sec:lep} and \ref{sec:magnetic} respectively. We obtain
\begin{align}
    \frac{\partial Y_{q_L^i}}{\partial z} =& \sqrt{\frac{90}{\pi^2 g_{\star}}} \frac{M_{\rm Pl}^{\prime}}{M_1}\left[-N_{\rm c} N_{\rm w} Q_q^2 \gamma_{\rm y} - N_c \mathcal{S}_{\rm w,sph} - N_{\rm w} \mathcal{S}_{\rm s,sph} - \sum_j \left( \mathcal{S}_{\rm uhu}^{ij} + \mathcal{S}_{\rm dhd}^{ij} \right)\right], 
    \label{eq:abol1}\\
    \frac{\partial Y_{u_R^i}}{\partial z} =& \sqrt{\frac{90}{\pi^2 g_{\star}}} \frac{M_{\rm Pl}^{\prime}}{M_1}\left[N_{\rm c} Q_u^2 \gamma_{\rm y} + \mathcal{S}_{\rm s,sph} + \sum_j \mathcal{S}_{\rm uhu}^{ji}\right], 
    \label{eq:abol3}\\
    \frac{\partial Y_{d_R^i}}{\partial z} =& \sqrt{\frac{90}{\pi^2 g_{\star}}} \frac{M_{\rm Pl}^{\prime}}{M_1}\left[N_{\rm c} Q_d^2 \gamma_{\rm y} + \mathcal{S}_{\rm s,sph} + \sum_j \mathcal{S}_{\rm dhd}^{ji}\right], 
    \label{eq:abol4}\\
    \frac{\partial Y_{l_L^i}}{\partial z} =& \sqrt{\frac{90}{\pi^2 g_{\star}}} \frac{M_{\rm Pl}^{\prime}}{M_1}\left[-N_{\rm w} Q_l^2 \gamma_{\rm y} - \mathcal{S}_{\rm w,sph} - \sum_j \mathcal{S}_{\rm Lhe}^{ij}\right] +\frac{1}{3}\Big[\epsilon_1 D_1 \left( Y_{N_1} - Y_{N_1}^{\rm  eq} \right) - W_1\sum_{j} Y_{l_L^j}\Big],
    \label{eq:abol2}\\
    \frac{\partial Y_{e_R^i}}{\partial z} =& \sqrt{\frac{90}{\pi^2 g_{\star}}} \frac{M_{\rm Pl}^{\prime}}{M_1}\left[Q_e^2 \gamma_{\rm y} + \sum_{j} \mathcal{S}_{\rm Lhe}^{ji}\right], 
    \label{eq:abol5}\\
    \frac{\partial Y_{\varphi}}{\partial z} =& \sqrt{\frac{90}{\pi^2 g_{\star}}} \frac{M_{\rm Pl}^{\prime}}{M_1}\left[\sum_{i,j} \left( \mathcal{S}_{\rm dhd}^{ij} - \mathcal{S}_{\rm uhu}^{ij} + \mathcal{S}_{\rm Lhe}^{ij} \right)\right]+\Big[\epsilon_1 D_1 \left( Y_{N_1} - Y_{N_1}^{\rm  eq} \right) - W_1\sum_{j} Y_{l_L^j}\Big],
    \label{eq:abol6}\\
    \frac{\partial Y_{N_1}}{\partial z} =& - D_1 \left( Y_{N_1} - Y_{N_1}^{\rm  eq} \right)\label{eq:abol7}.
\end{align}
Note that turning off the source term for the decay of right-handed neutrinos gives us baryogenesis through the hypermagnetic field, while setting $\gamma_{y}=0$ results in baryogenesis through thermal leptogenesis. In particular, if we add Eqs.\ (\ref{eq:abol1})-(\ref{eq:abol4}) and then subtract Eqs.\ (\ref{eq:abol5}) and (\ref{eq:abol6}), we obtain the evolution equation for $Y_{B-L}$ as given by Eq.~(\ref{eq:YBL}).

In order to formulate the equations presented earlier, we need to take into consideration the following key points:
First, in order to consider thermal leptogenesis, we make the assumption that the processes described in Eqs.\ (\ref{eq:decay}) and (\ref{eq:decay-anti}) produce an equal number of leptons in all three flavors regardless of the generation. Although using flavor leptogenesis would provide more precise results, we neglect it in this study to concentrate on the impact of the primordial magnetic field on leptogenesis.
Second, when RHNs decay, Higgs and leptons are produced equally. Hence, each decay of RHN contributes a factor of $\frac{1}{3}$ as a source for each generation of lepton doublet, and a factor of 1 as a source for each Higgs doublet, as expressed in Eqs.\ (\ref{eq:abol2}) and (\ref{eq:abol6}), respectively.
Finally, we again use the hypercharge neutrality condition given by Eq.\ (\ref{eq:hypercharge}), and solve the set of evolution equations given in  Eqs.\ (\ref{eq:abol1})-(\ref{eq:abol6}) and Eqs.\ (\ref{eq:YN1}) and (\ref{eq:YBL}) simultaneously. We can then obtain the baryon asymmetry by
\begin{align}
    Y_B^{\rm Lepto.+ U_{Y}(1)} = \frac{1}{3} \sum_i (Y_{q_L^i}+Y_{u_R^i}+Y_{d_R^i}),
\end{align}
where the superscript `$\rm Lepto.+U_{Y}(1)$' denotes the combined scenario.

\section{Numerical solutions and results}
\label{sec:res}
Now, we shall show how the hypermagnetic fields can affect the thermal leptogenesis in the early Universe. In this section, we numerically solve the set of evolution equations simultaneously for each of the three model, {\it i.e.}, the leptogenesis, baryogenesis through the ${\rm U}_Y(1)$ anomaly, and the combined model, separately, starting at $T^{\rm init}=10^{13}\ {\rm GeV}$ and ending at the $T_{\rm EWPT} = 100\ {\rm GeV}$. To have only leptogenesis mechanism active, we set $\gamma_{y}=0$ in Eqs. (\ref{eq:abol1}-\ref{eq:abol6}). Similarly, to have only baryogenesis through ${\rm U}_Y(1)$ anomaly, we disable the source term arising from the decay of right-handed neutrinos (RHNs). In the combined scenario, we solve the set of evolution equations, as presented in Eqs. (\ref{eq:abol1}-\ref{eq:abol7}), simultaneously. In the following, we set all of the initial Higgs, leptons and quarks asymmetries to zero. First of all, we must specify the Yukawa matrix given by Eq.\ (\ref{eq:Casas-Ibarra}). For the PMNS matrix, given by Eq.\ (\ref{eq:PMNS}), we take the standard neutrino oscillation parameters from NuFIT 5.2  \cite{Esteban:2020cvm}. For the $R$ matrix, given by Eq.\ (\ref{eq:R-matrix}), we choose the parameters shown in Tab.\ \ref{table:R-matrix}.
Moreover, as is stated earlier, we choose the second scenario of magnetic field evolution, as expressed in Eq.\ (\ref{eq:ICgy}), and set $B_{0}=10^{-14}\ \rm G$, $\lambda_0=10^{-6}\ \rm Mpc$, and $T_{\rm TS}=T_{\rm EWPT}=100\ \rm{GeV}$. Then, the helical source term given in Eq.\ (\ref{eq:ICgy}) reduces to
\begin{align}
    \gamma_{\rm y} 	\simeq   7.89 \times 10^{-24} \mathcal{C}.
    \label{eq:ICgy1}
\end{align}
Referring to Eqs.\ (\ref{gamma y}) and (\ref{eq:ICgy}), we can interpret the magnitude of coefficient $\mathcal{C}$ as a parameter for adjusting the strength of the initial hypermagnetic field, and $\mathrm{sgn}(\mathcal{C})=\mathrm{sgn}(\dot{h})$.
\begin{table}[h]
    \caption{R matrix parameter \label{table:R-matrix}}
    \begin{ruledtabular}
        \begin{tabular}{c c c c c c} 
            $x_1/\degree$ & $y_1/\degree$ & $x_2/\degree$ & $y_2/\degree$ & $x_3/\degree$ &   $y_3/\degree$ \\
            \colrule
            $180$ & $1.4$ & $180$ & $11.2$ & $180$ & $11$\\
        \end{tabular}
    \end{ruledtabular}
\end{table}

\begin{figure}[h]
    $M_2=M_1 \times 10^{0.6},\ M_3=M_1 \times 10^{1.0},\ Y_{N_1}^{\rm init}=0.004$
    \includegraphics[width=.9\textwidth]{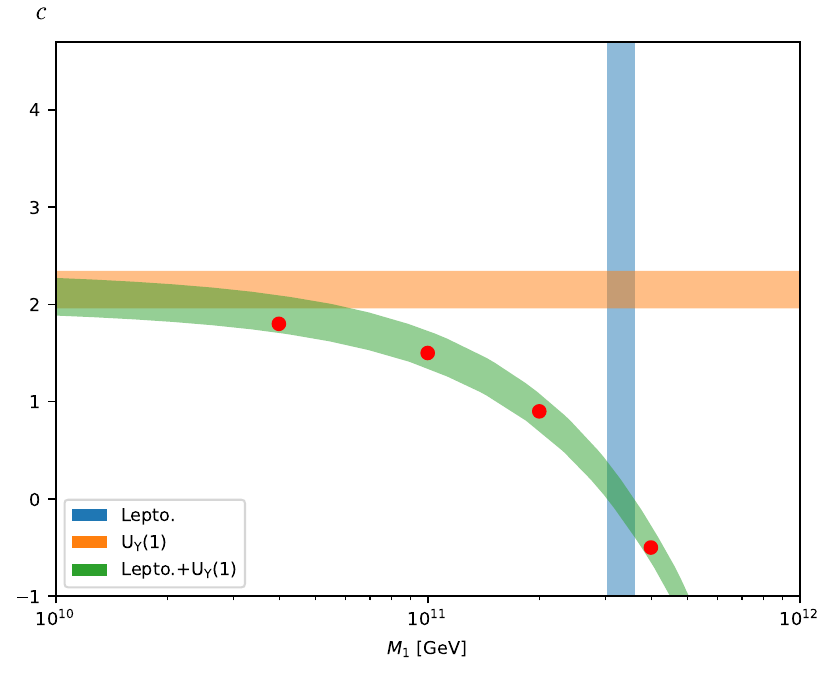}
    \caption{ Valid region of $\mathcal{C}$ and $M_1$ parameter space with $5\%$ deviation from $Y^{\rm obs}_{B}$ given by Eq.\ (\ref{eq:YBobs}). \label{fig:i}}
\end{figure}

First, we probe the $\mathcal{C}$ and $M_1$ parameter space for baryon asymmetry generated at the onset of EWPT time by the three $\rm Lepto.$, $\rm U_{Y}(1)$, and $\rm Lepto.+U_{Y}(1)$ scenarios. We solve the relevant evolution equations, for normal mass hierarchy, with $M_2=M_1\times 10^{0.6}\ \rm {GeV}$, $M_3=M_1\times 10^{1}\ \rm {GeV}$, and initial conditions $Y^{\rm  init}_{N_1} =0.004\neq Y^{\rm  eq}_{N_{1}}$\footnote{Note that by using Eq.\ (\ref{eq:YEq}) we have $Y_{N_1}^{\rm  eq}(z_0) = \frac{45}{4 \pi^4} \frac{g_{N_1}}{g_{\star}}z_{0}^2 K_2(z_0)\approxeq 0.004317$.}. The results are shown in Fig.\ \ref{fig:i}. 
Using the points in the green region in this figure results in values of $Y^{\rm Lepto.+U_{Y}(1)}_B$ at the EWPT which are within $5\%$ of $Y^{\rm obs}_{B}$. We shall henceforth refer to this region as the physically relevant region. Note that this region includes regions which are not included in the other two models. In fact, our model reduces the minimum values of both $M_1$ and $\mathcal{C}$ needed for generating the acceptable baryon asymmetry. In particular, the values of RHNs masses decrease by at least one order of magnitude.

\begin{figure}[h]
    $M_2=M_1 \times 10^{0.6},\ M_3=M_1 \times 10^{1.0},\ Y_{N_1}^{\rm init}=0.004$
    \includegraphics[width=.49\textwidth]{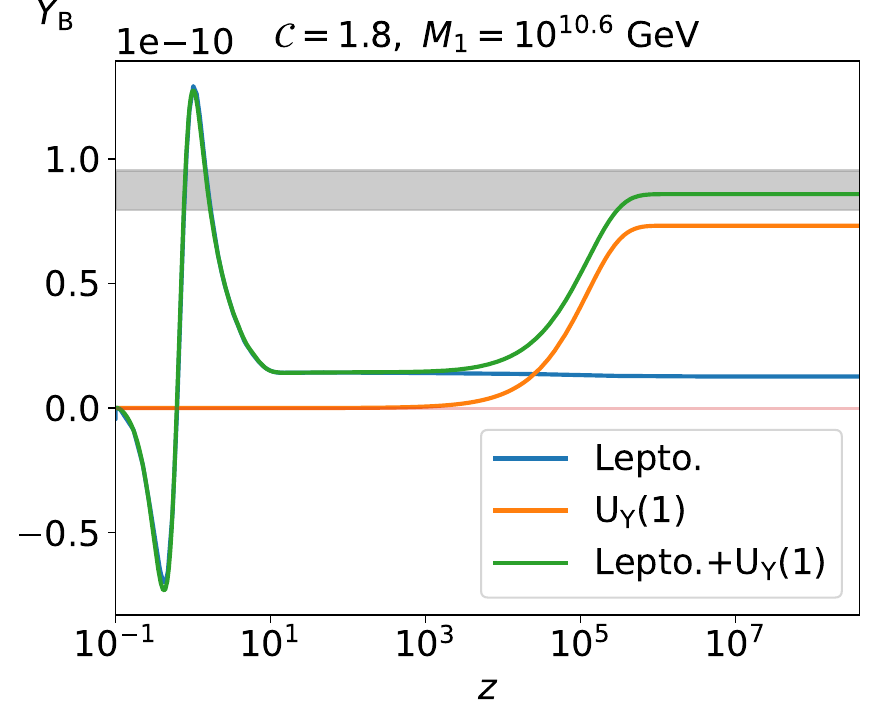}
    \hfill
    \includegraphics[width=.49\textwidth]{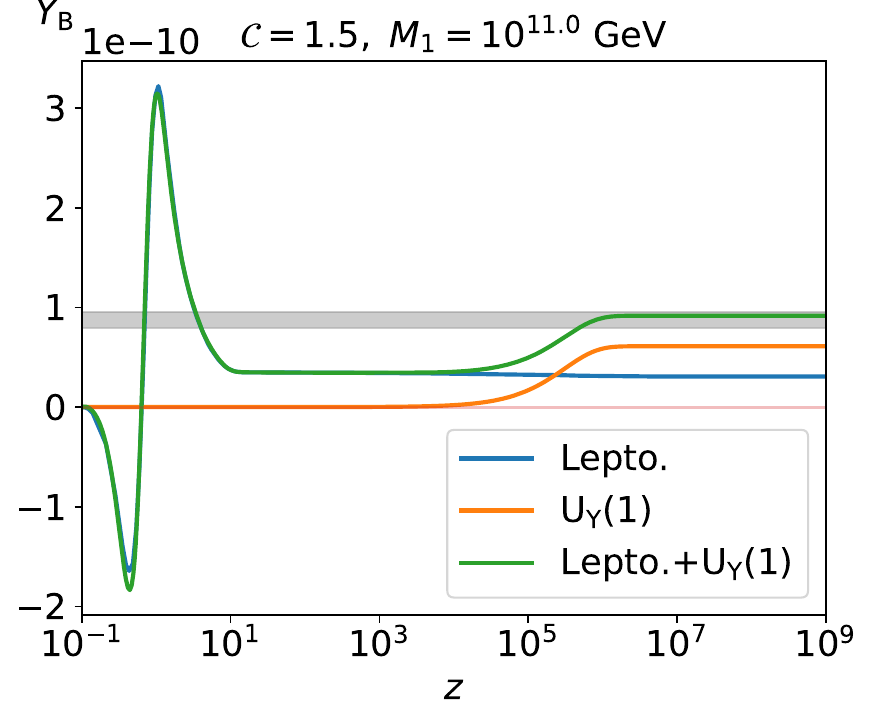}
    \hfill
    \includegraphics[width=.49\textwidth]{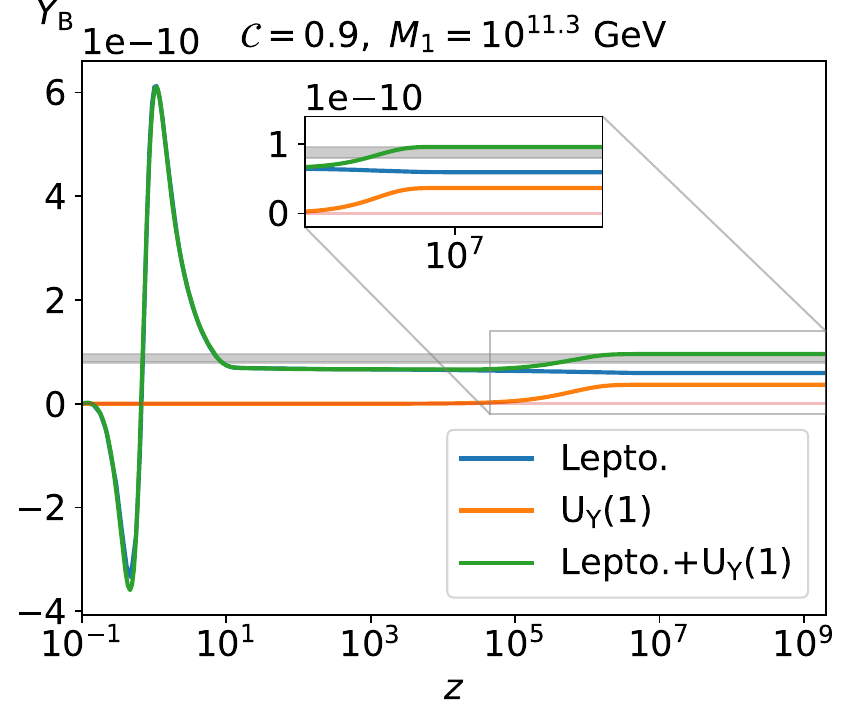}
    \hfill
    \includegraphics[width=.49\textwidth]{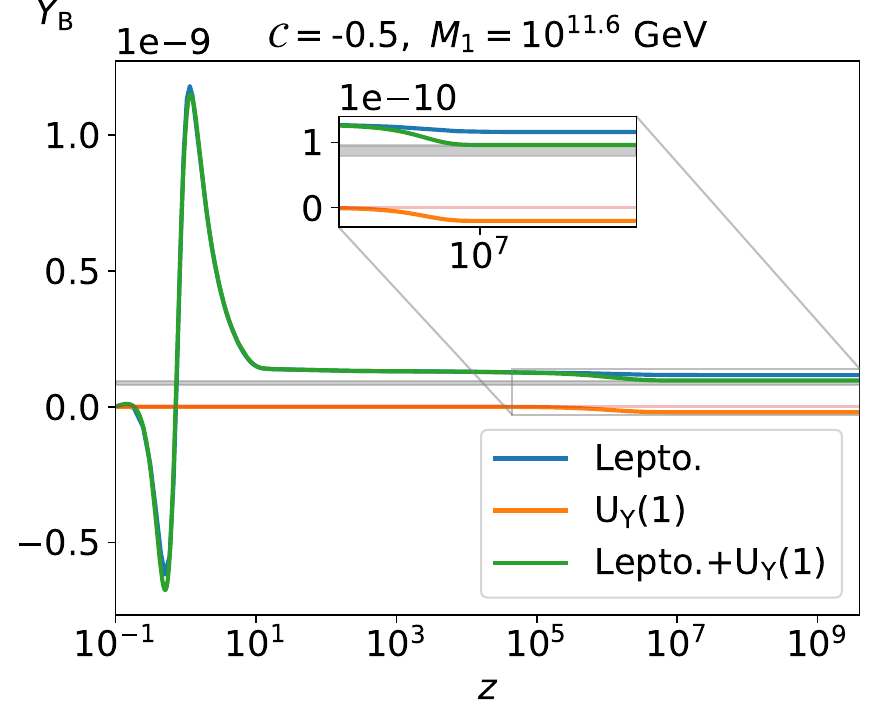}
    \caption{Time evolution of baryon asymmetry generated via $\rm Lepto.$, $\rm U_{Y}(1)$, and $\rm Lepto.+U_{Y}(1)$. The gray region shows $Y^{\rm obs}_{B}$ within $5\%$ deviation from Eq.\ (\ref{eq:YBobs}). \label{fig:ii}}
\end{figure}

Second, we numerically solve the evolution equations of $\rm Lepto.$, $\rm U_{Y}(1)$, and $\rm Lepto.+U_{Y}(1)$ scenarios for four points of parameter space, within the physically relevant region,  indicated in Fig.\ \ref{fig:i} by red dots, and present the results in Fig.\ \ref{fig:ii}. Our results show that for $\mathcal{C}>0$ ($\mathcal{C}<0$), $Y^{\rm U_{Y}(1)}_{B}$ is positive (negative) at the EWPT, whereas $Y^{\rm Lepto.}_B$ at EWPT is always positive. Moreover, we observe that for the cases shown in Fig.\ \ref{fig:ii}, $Y^{\rm Lepto.+U_{Y}(1)}_B\approxeq Y^{\rm Lepto.}_B + Y^{\rm U_{Y}(1)}_{B}$. However, we shall present a drastic and yet interesting counterexample below.

\begin{figure}[h]
    $M_1= 10^{11.4}\ {\rm GeV},\ M_2=10^{12.0}\ {\rm GeV},\ M_3=10^{12.4}\ {\rm GeV}, Y_{N_1}^{\rm init}=Y_{N_1}^{eq}$
    \includegraphics[width=.49\textwidth]{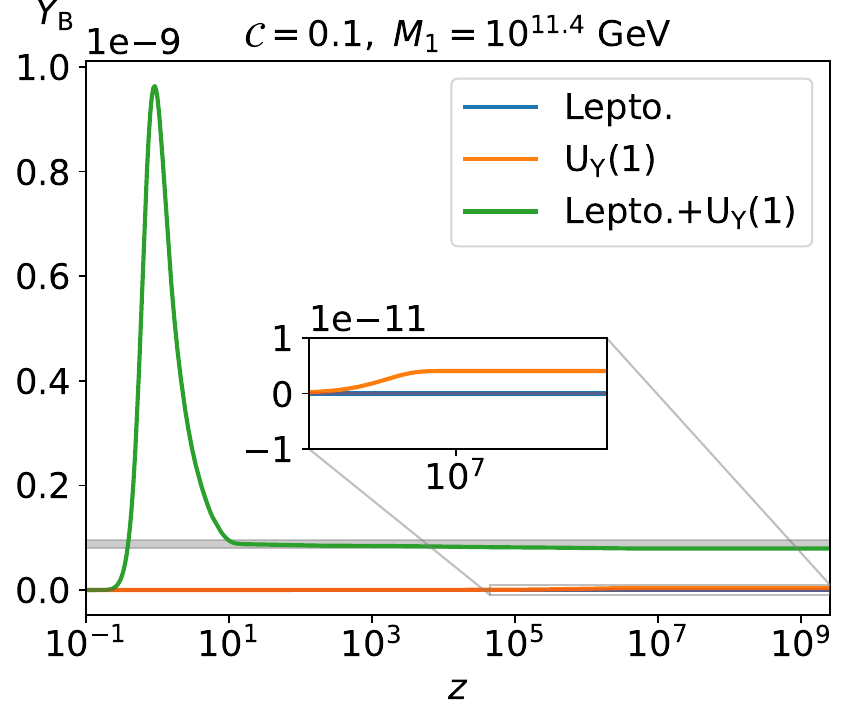}
    \hfill
    \includegraphics[width=.49\textwidth]{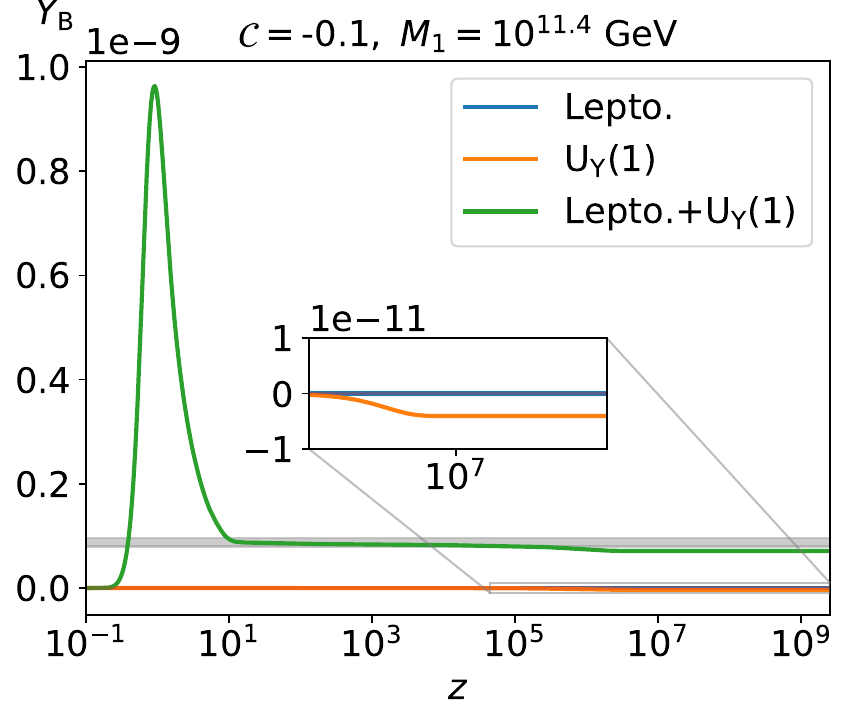}
    \caption{ime evolution of baryon asymmetry generated via $\rm Lepto.$, $\rm U_{Y}(1)$, and $\rm Lepto.+U_{Y}(1)$ with initial conditions mentioned in sub-figures' title. \label{fig:iii}}
\end{figure}

Finally we investigate the case $Y_{N_1}^{\rm init} = Y_{N_1}^{\rm eq}$, which can be be considered as a more natural initial condition. In Fig.\ \ref{fig:iii}, we display the results for $M_1 =10^{11.4}$ and $\mathcal{C} = \pm 0.1$, which is again a point in the physically relevant region. As can be seen from this figure, $Y^{\rm Lepto.}_B=0$, as expected, while $Y^{\rm U_{Y}(1)}_{B}$ at the EWPT is approximately $\pm 4\times 10^{-12}$. However, interestingly, $Y^{\rm Lepto.+U_{Y}(1)}_B \approxeq 10^{-10}$ at the EWPT, independent of sign of $\mathcal{C}$ and hence that of $Y^{\rm U_{Y}(1)}_{B}$.

The reason for this outcome can be explained as follows: In pure thermal leptogenesis, when we set $Y_{N_1}^{\rm init} = Y_{N_1}^{\rm eq}$, the term proportional to $\epsilon_1 D_1 \left( Y_{N_1} - Y_{N_1}^{\rm eq} \right)$ in Eq.\ (\ref{eq:YBL}) becomes zero. Consequently, $B-L$ freezes at zero because there are no initial asymmetries. Therefore, the baryon asymmetry is not generated through the thermal leptogenesis, $Y^{\rm Lepto.}_B=0$. However, $Y^{\rm U_{Y}(1)}_{B}$ can be produced through the hypermagnetic helicity decay. Therefore, $Y^{\rm Lepto.}_B+Y^{\rm U_{Y}(1)}_B=Y^{\rm U_{Y}(1)}_{B}$ in a scenario where we consider thermal leptogenesis and baryogenesis through hypermagnetic helicity separately. Now, in a combined scenario, even though the term involving $\epsilon_1 D_1 \left( Y_{N_1} - Y_{N_1}^{\rm eq} \right)$ becomes zero for $Y_{N_1}^{\rm init} = Y_{N_1}^{\rm eq}$, the washout term $W_1 Y_{l_L}$ can still be active after lepton asymmetry generation through hypermagnetic helicity decay. Consequently, in this combined scenario, thermal leptogenesis can generate a nonzero $B-L$ asymmetry and as a result a non zero baryon asymmetry, unlike the separated scenario. So, $Y^{\rm Lepto.+U_{Y}(1)}_B$ is not equal to $Y^{\rm U_{Y}(1)}_B$.

\section{Conclusion}
\label{sec:conclu}
In this work, we have investigated the production and evolution of the baryon asymmetry through thermal leptogenesis in the presence of nonzero background helical hypermagnetic fields. 
To this end, we have combined two models: baryogenesis through ${\rm U}_Y(1)$ anomaly and thermal leptogenesis. To write the evolution equations, we started with those of the former and added the evolution equation of the RHN. Moreover, the evolution equations of the left-handed leptons and Higss asymmetries acquire two additional sources: the out-of-equilibrium decay of the lightest RHN and the time evolution of the hypermagnetic helicity and their washout terms. In this work, we have considered the hypermagnetic fields that initially evolve adiabatically, then enter the inverse cascade regime at transition temperature $T=T_{\rm EWPT}$. We have assumed that the asymmetries produced through the out-of-equilibrium decay of the lightest RHN are distributed equally between the Higgs and leptonic parts. That is, we have neglected the flavor effects in the leptogenesis scenario and focused on the vanilla scenario with equal distribution for three lepton flavors. We have shown that the physically relevant region of the parameter space in our combined model differs from those of the individual models considered separately. In particular, we have found that in the presence of a nonzero background hypermagnetic field, the mass scale of the RHNs can be decreased by at least one order of magnitude. 

We have shown that the baryon asymmetry generated at the EWPT in the combined scenario is not always approximately equal to the sum of the values generated by its two components separately.
In particular, we have found that the combined scenario can produce the desired baryon asymmetry in the presence of weak hypermagnetic fields, even if the initial abundance of the RHN is exactly at its equilibrium value. This is unlike the common leptogenesis scenario were an initial deviation from the equilibrium abundance is necessary to produce the desired BAU, and indicates a significant synergy.

\acknowledgments
The authors are very grateful to S. S. Gousheh for his valuable comments, assistance, and proofreading of the manuscript.
SA acknowledges the financial support from the Iran National Science Foundation (INSF) through grant No. 4003903.
SA also acknowledges support by the European Union’s Framework Programme for Research and Innovation Horizon 2020 under the Marie Sklodowska-Curie grant agreement No 860881-HIDDeN as well as under the Marie Sklodowska-Curie Staff Exchange grant agreement No 101086085-ASYMMETRY.
SA would like to thank Shahid Beheshti University for financial support.

\bibliography{biblio}

\end{document}